\documentclass[modern]{aastex61}

\usepackage{natbib}
\usepackage[utf8]{inputenc}
\usepackage{graphicx}
\usepackage{amsmath}
\usepackage{hyperref}

\newcommand{\Romer}{R{\o}mer}

\usepackage{color}
\usepackage[normalem]{ulem}

\shorttitle{Barycentric and Asymmetric Transverse Velocities}
\shortauthors{Conroy et al.}

\begin{document}

\title{The Effects of Barycentric and Asymmetric Transverse Velocities on Eclipse and Transit Times}

\author[0000-0002-5442-8550]{Kyle E.~Conroy}
\affiliation{Vanderbilt University, Dept.~of Physics and Astronomy, 6301 Stevenson Center Ln, Nashville TN, 37235, USA}
\affiliation{Villanova University, Dept.~of Astrophysics and Planetary Sciences, 800 E.\ Lancaster Ave, Villanova, PA 19085, USA}

\author{Andrej Pr\v sa}
\affiliation{Villanova University, Dept.~of Astrophysics and Planetary Sciences, 800 E.\ Lancaster Ave, Villanova, PA 19085, USA}

\author{Martin Horvat}
\affiliation{University of Ljubljana, Dept.~of Physics, Jadranska 19, SI-1000 Ljubljana, Slovenia}
\affiliation{Villanova University, Dept.~of Astrophysics and Planetary Sciences, 800 E.\ Lancaster Ave, Villanova, PA 19085, USA}

\author{Keivan G.~Stassun}
\affiliation{Vanderbilt University, Dept.~of Physics and Astronomy, 6301 Stevenson Center Ln, Nashville TN, 37235, USA}
\affiliation{Fisk University, Department of Physics, 1000 17th Ave.\ N., Nashville, TN 37208, USA}


\begin{abstract}

It has long been recognized that the finite speed of light can affect the observed time of an event.  For example, as a source moves {\it radially} toward or away from an observer, the path length and therefore the light travel time to the observer decreases or increases, causing the event to appear earlier or later than otherwise expected, respectively.  This light travel time effect (LTTE) has been applied to transits and eclipses for a variety of purposes, including studies of eclipse timing variations (ETVs) and transit timing variations (TTVs) that reveal the presence of additional bodies in the system.  
Here we highlight another non-relativistic effect on eclipse or transit times arising from the finite speed of light---caused by an asymmetry in the {\it transverse} velocity of the two eclipsing objects, relative to the observer.  This asymmetry can be due to a non-unity mass ratio or to the presence of external barycentric motion.  Although usually constant, this barycentric and asymmetric transverse velocities (BATV) effect can vary between sequential eclipses if either the path length between the two objects or the barycentric transverse velocity varies in time.  
We discuss this BATV effect and estimate its magnitude for both time-dependent and time-independent cases. 
For the time-dependent cases, we consider binaries that experience a change in orbital inclination, eccentric systems with and without apsidal motion, and hierarchical triple systems.
We also consider the time-independent case which, by affecting the primary and secondary eclipses differently, can influence the inferred system parameters, such as the orbital eccentricity.

\end{abstract}

\section{Introduction}

The so-called \emph{\Romer\ delay}, named after Ole \Romer\ who computed the speed of light from the eclipses of Io by Jupiter in 1676, has long been applied to eclipsing binary star systems and transiting planets to account for the effect of the finite speed of light on the observed timings of eclipses. \citet{Borkovits2003, Borkovits2007, Borkovits2011} provide analytical expressions for both the LTTE and dynamical effects of a third body on the eclipse timings of an inner-binary.  This LTTE effect, focuses on the change in the distance that light must travel to reach the observer as the inner-binary orbits around the common center of mass of the entire system, whereas the dynamical effects deal with perturbations to the orbital elements (predominantly the orbital period) of the inner-binary caused by interactions between the orbits. 

\citet{Kaplan2010} used the finite speed of light, along with the variation of the photon path length relative to the barycenter as a function of mass-ratio, to compute the mass-ratio of a given system by precisely measuring the eclipse times.  This has successfully been applied to observable systems, including a \emph{Kepler} eclipsing sdB+dM binary \citep{Barlow2012}.  Similarly, \citet{Loeb2005} accounted for the finite speed of light and the fact that, near eclipse or transit, the line-of-sight projected velocity of the eclipsing body changes sign, resulting in an asymmetry between the ingress and egress slopes of the eclipse \citep[see also][]{Barnes2007}.  

\citet{Shklovskii1970} accounted for the transverse velocity of pulsars resulting in a positive time-derivative to the observed rotation period.  \citet{Kaplan2014} applied this to eclipse times in the case of a double white-dwarf binary.  \citet{Scharf2007} and \citet{Rafikov2009} discuss the effects of parallax and proper motion on transit times due to the apparent precession of the orbit as the system moves on the plane of sky.  Similarly, \citet{Scharf2007} also discusses the resulting change in transit duration due to an apparent change in inclination.

As the precision of observed eclipse timings improves, it becomes increasingly important to account for higher-order effects.  Eclipse timings were measured in bulk for all detected eclipsing binaries in the \emph{Kepler} long cadence (30 min) data-set to a precision on the order of tens of seconds to minutes, depending on the stellar noise and orbital period of the system \citep{Gies2012, Rappaport2013, Conroy2014, Orosz2015, Borkovits2015, Borkovits2016}.  For systems with high-precision follow-up or manual fitting, individual eclipse times with uncertainties on the order of seconds are not unreasonable.

Here we present a non-relativistic contribution to the observed timings of eclipses (hereafter, our use of the word ``eclipse'' can also be applied to occultations and transits) caused by any asymmetry in the transverse velocities of the two objects relative to the observer.  This can be the result of internal (i.e.~non-unity mass-ratio) or external (i.e.~barycentric motion, additional components) causes.  GAIA Data Release 1 (DR1) has recently released proper motions for 2 million sources brighter than 20.7 magnitude \citep{gaia2016a, gaia2016b, Lindegren2016}, making it possible to obtain this information for any source in the very near future.

In Section \ref{sec:theory}, we provide the formalism describing this effect, hereafter BATV for ``Barycentric and Asymmetric Transverse Velocities''.  In Section \ref{sec:gaia}, we use parallaxes and proper motions from GAIA DR1 to obtain the expected distribution in transverse velocities that in turn influence the expected magnitude of the BATV signal.  In Section \ref{sec:scenarios}, we then examine several physical scenarios in which BATV plays a role and also discuss the consequence of a time-dependent change in the radial separation of the two components between successive eclipses.  In many of these cases, the contribution to the eclipse times from BATV is negligible or unlikely to be observable, but we show that failing to account for the contribution could result in erroneous measurements of orbital or physical parameters. We summarize our conclusions in Section~\ref{sec:conclusion}.

\section{General Theory}\label{sec:theory}

In order to define the observed time of eclipse, the positions of both bodies must be individually corrected according to the light travel time between their respective instantaneous positions and some fixed reference frame.  Any asymmetry in the transverse velocities of these two bodies relative to the observer, therefore, results in an unequal correction in their positions.  This asymmetry can result from two root causes.  A non-unity mass-ratio in the system will result in the lower-mass object having a higher velocity than the higher-mass object at any given time throughout the orbit.  Additionally, any barycentric transverse (i.e.~on the plane of the sky rather than along the line of sight) motion will affect the absolute transverse velocities of both components, relative to the observer.  Here we provide a derivation of the effect that BATV has on the observed time of eclipse.

\begin{figure}[b!]
    \centering
    \includegraphics[width=\textwidth]{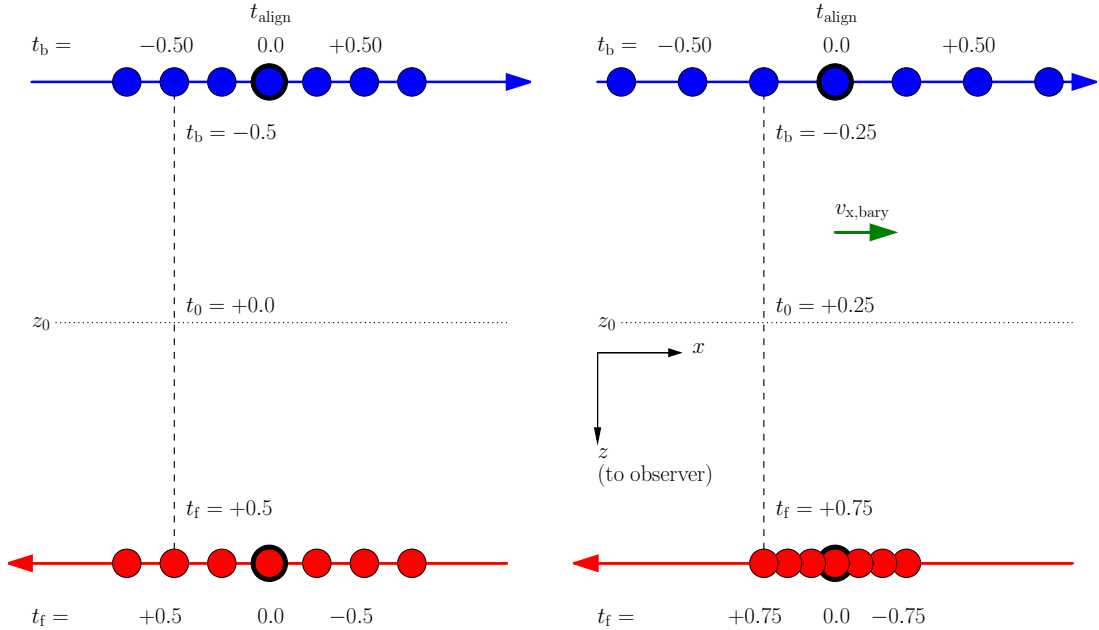}
    \caption{2D schematic representation of the effect BATV has on observed eclipse times for an equal-mass binary system.  On the left is a system with no barycentric motion, such that both components have equal---but opposite---velocities.  On the right is a representation of the same system, but with the addition of transverse barycentric motion, such that the speed of the star in front is $1/3$ the speed of the star in back, relative to the observer.  The separation between the two stars in both cases is equivalent to one light-time unit.  At time $t=t_\text{align}=0.0$, the stars are in geometric alignment (i.e.~the time of eclipse as provided by the ephemeris as $c \to \infty$).  An eclipse is observed when a photon that was emitted by the back (blue) star travels the distance to the front (red) star and is intercepted.  In other words, the x-position of the stars must align while being separated by exactly the time it takes the photon to cross the distance between them. We define the time of eclipse as the time at which the photon passes the plane containing the system barycenter, $z_0$.  For the case with no barycentric motion (left), the eclipse is observed at t=0 (although shifted to the left in space), whereas the case with positive barycentric motion (right) has a shift in the observed eclipse time by 0.25 light-time units.}
    \label{fig:schematic}
\end{figure}

If we assume that all orbital motion is in the $xz$ plane with $z$ pointing towards the observer (therefore guaranteeing an eclipse), then we can define the condition for an observed eclipse as:
\begin{equation}
    x_\text{b} (t_\text{b}) = x_\text{f} (t_\text{f}) \>,
\end{equation}
where the subscripts ``b'' and ``f'' refer to the back (eclipsed) and front (eclipsing) star, respectively.  This condition states that the stars must appear aligned w.r.t.~a photon traveling towards the observer.  The photon that is emitted by the back star at $x_\text{b} (t_\text{b})$ at time $t_\text{b}$ will be intercepted by the front star at $x_\text{f} (t_\text{f})$ at time $t_\text{f}$ (see Figure \ref{fig:schematic} for a schematic).  By expressing the x-positions of both stars in terms of their x-velocities, our condition for an observed eclipse becomes:
\begin{equation}\label{eq:condition_eclipse}
    - \int_{t_\text{b}}^{t_\text{align}} v_\text{x,b}(t) dt = \int_{t_\text{align}}^{t_\text{f}} v_\text{x,f}(t) dt \>,
\end{equation}
where $t_\text{align}$ is the time at which the stars are in geometric alignment, i.e.~when $c \to \infty$.

For any photon traveling in the positive z-direction towards the observer between the two stars, the times must satisfy the following condition, accounting for the light travel time:
\begin{equation}\label{eq:condition_ltte}
    t_\text{f} - t_\text{b} = \frac{z_\text{f} (t_\text{f}) - z_\text{b} (t_\text{b})}{c} \>.
\end{equation}

The time of observed eclipse can be given w.r.t.~the photon crossing any plane of choice along $z$; a convenient choice which also allows using this time shift in conjunction with LTTE is the plane that contains the barycenter of the system, $z_0$. By making this choice, the resulting expression can be used in conjunction with classical LTTE which accounts for the shift due to the travel time between the barycenter and the observer. We can express the time, $t_0$, at which the photon crosses this $z_0$ plane as follows:
\begin{equation}\label{eq:condition_t0}
    t_0 = t_\text{f} - \frac{z_\text{f} (t_\text{f}) - z_0 (t_0)}{c} \>.
\end{equation}

To find the time of observed eclipse, we need to solve Equations (\ref{eq:condition_eclipse}), (\ref{eq:condition_ltte}), and (\ref{eq:condition_t0}) for $t_0$.  If the functional dependence of $v_\text{x,b}(t)$ and $v_\text{x,f}(t)$ is known, this can be computed either analytically or numerically. For the purposes of deriving an approximate general analytic solution (see Appendix \ref{app:approximation} for an estimate on the error introduced by this approximation), let us examine the case where the x-velocities can be assumed constant throughout the travel time of the photon, $[t_\text{b}, t_\text{f}]$, thereby allowing us to simplify Equation (\ref{eq:condition_eclipse}) as follows:
\begin{equation}\label{eq:condition_eclipse_approx}
    \left( t_\text{b} - t_\text{align} \right) v_\text{x,b}  = \left(t_\text{f} - t_\text{align} \right) v_\text{x,f}  \>.
\end{equation}

We can now use Equations (\ref{eq:condition_ltte}), (\ref{eq:condition_t0}), and (\ref{eq:condition_eclipse_approx}) to solve for $\Delta t_\text{BATV} \equiv t_0 - t_\text{align}$, i.e.~the time {\it shift}, relative to the time of geometric alignment, at which the photon emitted by one star, traveling along the line of sight, and then intercepted by another star, will pass the $z_0$ plane:
\begin{equation}\label{eq:general}
\begin{split}
    \Delta t_\text{BATV} &= \frac{ z_\text{f} (t_\text{f}) - z_\text{b} (t_\text{b})}{c} \frac{ v_\text{x,b} }{ v_\text{x,b} - v_\text{x,f} } - \frac{z_\text{f} (t_\text{f}) - z_0 (t_0)}{c} \>.
\end{split}
\end{equation}

If the z-positions of both stars and the barycenter are constant (see Appendix \ref{app:binary_linear} for a discussion that instead uses a linear approximation) over the photon path time interval, $[t_\text{b}, t_\text{f}]$, such that $z_\text{f} (t_\text{f}) = z_\text{f} (t_0)$ and $z_\text{b} (t_\text{b}) = z_\text{b} (t_0)$, then we can simplify Equation (\ref{eq:general}) by dropping all dependencies on time as follows:
\begin{equation}\label{eq:general_constant_z}
    \Delta t_\text{BATV} = \frac{ \Delta z_\text{bf} }{c} \frac{ v_\text{x,b} }{ v_\text{x,b} - v_\text{x,f} } - \frac{ \Delta z_\text{0f} }{ c } \>,
\end{equation}
where $\Delta z_\text{bf} \equiv z_\text{f} - z_\text{b}$ and $\Delta z_\text{0f} \equiv z_\text{f} - z_0$.  Note again that all values of $\Delta z_{ij}$ and $v_{x,i}$ may change between successive eclipses (see Section \ref{sec:scenarios} for example cases), but are assumed constant over the light travel time between the two stars at eclipse.  

In order to determine the observed time of any individual eclipse, this effect, as well as any delay caused by a change in the distance between the observer and the barycenter (i.e.~classical LTTE) must be taken into account:
\begin{equation}
    t_\text{obs} = t_\text{align} + \Delta t_\text{LTTE} + \Delta t_\text{BATV} \>,
\end{equation}
where $t_\text{align}$ itself may need a dynamical correction for any perturbations to the orbital period or other elements (e.g.~due to interactions with additional bodies in the system) from the value provided by a linear ephemeris: 

\begin{equation}
    t_\text{align} = t_\text{ephem} + \Delta t_\text{dyn} \>.
\end{equation}

\subsection{Application to Keplerian Orbits}

In the case of a Keplerian binary system, we can further simplify by expressing the z-position of the barycenter, $z_0$, in terms of the mass-ratio of the binary.  From the definition of the center of mass, we know that $\Delta z_{0f} =   \Delta z_\text{bf} / \left( \xi+1 \right)$ where $\xi \equiv M_\text{f} / M_\text{b}$ is the mass-ratio, $q$, for a primary eclipse, or the inverse of the mass-ratio, $1/q$, for a secondary eclipse, giving:

\begin{equation}\label{eq:general_binary}
    \Delta t_\text{BATV} = \frac{\Delta z_\text{bf}}{c} \left( \frac{v_\text{x,b}}{v_\text{x,b} - v_\text{x,f}} - \frac{1}{\xi+1} \right) \>.
\end{equation}

We can separate the external barycentric velocity (denoted with the `bary' subscript) and orbital velocities relative to that same barycenter (denoted with the `orb' subscript), and take advantage of the relationship that, for a Keplerian orbit, $\xi = M_\text{f} / M_\text{b} = | \bf{v_\text{b,orb}} | / | \bf{v_\text{f,orb}} | $.  So, by substituting $v_\text{x,f} = v_\text{x,f,orb} + v_\text{x,bary}$ and $v_\text{x,b} = - \xi v_\text{x,f,orb} + v_\text{x,bary}$, we get the following final expression for the time shift of an eclipse caused by BATV:

\begin{equation}\label{eq:general_binary_vext}
    \Delta t_\text{BATV} = \frac{\Delta z_\text{bf}}{c} \left( \frac{\xi-1}{\xi+1} - \frac{1}{\xi + 1}\frac{v_\text{x,bary}}{v_\text{x,f,orb}} \right) \>.
\end{equation}

The same derivation but with linear (instead of fixed) motion along the line-of-sight is shown in Appendix \ref{app:binary_linear}, and the orbital velocity ($v_\text{x,f,orb}$) and separation ($\Delta z_\text{bf}$) terms are provided as orbital elements in Appendix \ref{app:binary}.

In many cases, the shift in observed eclipse times is constant and can simply be absorbed by a time-offset in the entire light curve.  However, if any of the above quantities vary with time, then this shift, $\Delta t_\text{BATV}$, also varies in time, resulting in a contribution to the ETVs.

\section{Distribution of Transverse Velocities in GAIA}\label{sec:gaia}

GAIA DR1 \citep{gaia2016a, gaia2016b, Lindegren2016} includes measured parallaxes and proper motions for $\sim$ 2 million sources also found in the  Hipparcos and Tycho-2 catalogs.  With approximately 1 billion total targets in GAIA, the number of sources with precisely determined proper motions can be expected to increase drastically in the near future.  As BATV depends strongly on the transverse velocity of a system, GAIA will enable us to estimate the magnitude of BATV for most observed systems.

Unfortunately, proper motions alone are not enough, as BATV depends on these transverse velocities projected along the direction of motion of the eclipsed object on the sky.  In some rare cases, this orientation of a given system on the sky may be constrained, e.g.~through direct imaging or astrometric solutions, but in most cases it will likely remain unknown.

\begin{figure}[bth!]
    \centering
    \includegraphics[width=\textwidth]{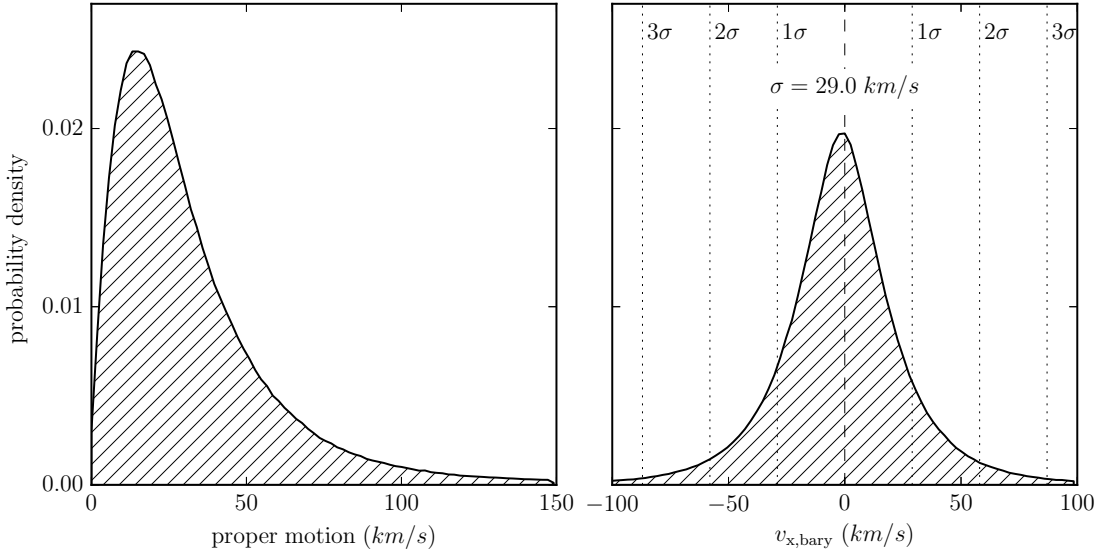}
    \caption{Left: the distribution of proper motions (in velocity units) computed from the GAIA DR1 (excluding the 3\% of targets with proper motions above $150~km s^{-1}$).  Right: the derived distribution in $v_\text{x,bary}$, where the velocity is determined from a randomized orientation of the binary on the sky relative to the proper motion. The standard deviation ($29~km s^{-1}$) and the corresponding confidence levels are shown with vertical dotted lines.}
    \label{fig:gaia_dist}
\end{figure}

The existing proper motions from GAIA DR1, however, provide the expected distributions of the projected transverse velocity, $v_\text{x, bary}$.  Figure \ref{fig:gaia_dist} depicts the distribution of proper motions (in velocity units) computed directly from the GAIA parallaxes and proper motions as well as the distribution of projected transverse velocities, assuming a random distribution in orientations of binary systems on the sky.  Excluding the 3\% of sources with proper motions above $150~km s^{-1}$, the expected projected transverse velocity is up to $100~km s^{-1}$, with 68\% ($1\sigma)$ falling between $-29$ and $+29~km s^{-1}$, 95\% ($2\sigma$) falling between $-58$ and $+58~km s^{-1}$, and 99.7\% ($3\sigma$) falling between $-87$ and $+87~km s^{-1}$.  Although these proper motions alone will not allow for estimating the exact value of $v_\text{x, bary}$ for a particular system, it does allow an estimate for a statistical range of values.

\section{Specific Cases}\label{sec:scenarios}

In Section \ref{sec:scenarios_phasesep} we discuss the constant, time-independent shift in eclipse times.  When this affects the primary and secondary eclipses to the same extent, the resulting shift in observed eclipse times is not noticeable and will be absorbed by an apparent time- or phase-shift.  However, in some cases the primary and secondary eclipses are affected differently, resulting in a constant shift in the observed {\it phase-separation} between the primary and secondary eclipses.

In Sections \ref{sec:scenarios_apsidal}-\ref{sec:scenarios_triple} we discuss several cases in which there is a time-dependent effect on the observed eclipse times.  These variations can be directly observed (given a sufficiently large magnitude) in addition to classical LTTE and dynamical effects by measuring the observed times of eclipses and comparing to the linear ephemeris.  Generally speaking, BATV becomes time-dependent whenever there is a time-dependent change in the separation between the components $\Delta z_\text{bf}$ (i.e.~for apsidal motion or a change in inclination) or in the barycentric transverse velocity $v_\text{x,bary}$ (i.e.~for the inner-binary in a hierarchical system) between successive eclipses.

Note that, for simplicity, all cases below will use Equation (\ref{eq:general_binary_vext}) which makes the following assumptions: the transverse velocities ($v_{x,*}$) and radial positions of both stars and the system as a whole are constant throughout the light travel time interval between the two stars.  As is estimated in Appendix \ref{app:approximation}, this approximation introduces more errors with an increase in the travel time between the two components and an increase in the velocities of the components.

\subsection{Constant Shift in Phase-Separation Between Eclipses}\label{sec:scenarios_phasesep}

\begin{figure}[bth!]
    \centering
    \includegraphics[width=\textwidth]{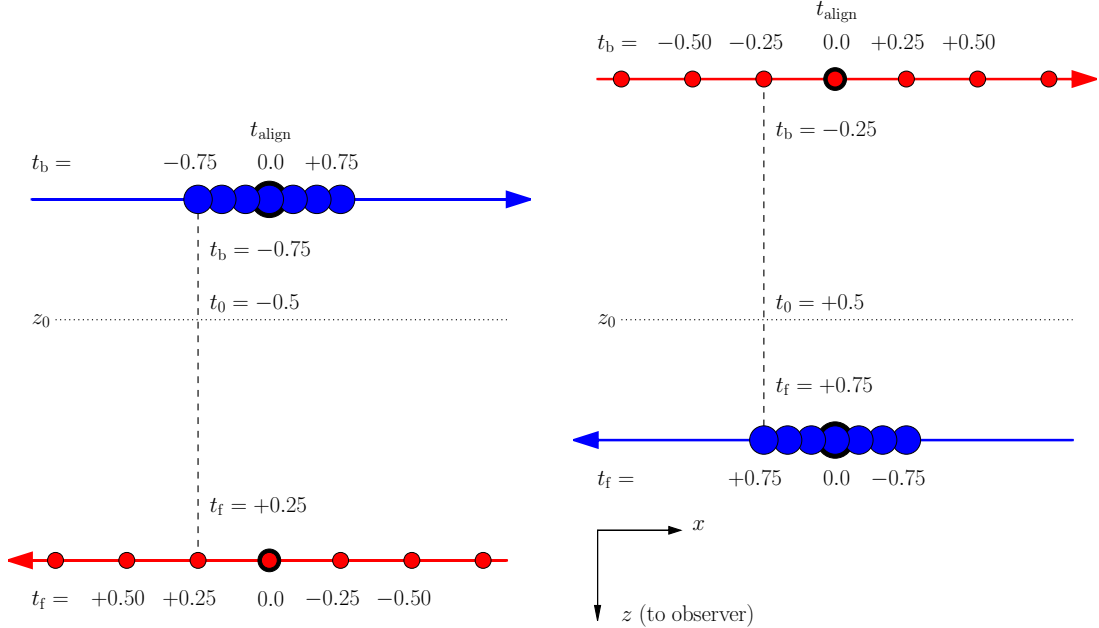}
    \caption{2D schematic representation showing how BATV with non-unity mass-ratio ($q=1/3$ in the case shown) affects the phase-separation between primary and secondary eclipses.  Left: a primary eclipse ($\xi \equiv q = 1/3$), with the more massive object (with a slower velocity and closer to the barycenter) being eclipsed.  Right: a secondary eclipse ($\xi \equiv 1/q = 3$) with the roles reversed.  Since the observed time is measured with respect to the (fixed) barycenter, the primary and secondary eclipses are observed to be shifted with respect to each other.}
    \label{fig:schematic_q}
\end{figure}

For a binary system with time-independent (or zero) barycentric transverse velocity, the observed times of the primary and secondary eclipses, relative to each other, can still be altered for a non-equal mass system due to both the asymmetric velocities and the distance from the barycenter during eclipse, as depicted schematically in Figure \ref{fig:schematic_q}.  This same effect, for the case without barycentric velocity, was discussed in \citet{Kaplan2010} and \citet{Fabrycky2010}.

We can express the magnitude of this effect as the difference between the time shifts for the primary and secondary eclipses, $t_\text{pri}$ and $t_\text{sec}$, divided by the orbital period, $P$, used for phasing.  This resulting $\Delta \Phi_\text{sep, BATV}$ will be the observed change in phase-separation between the primary and secondary eclipses as compared to the expected value (i.e.~0.5 for a circular system, assuming $c \to \infty$).

Here we make the assumption that $\Delta z_\text{bf}$ (provided in terms of orbital elements in Appendix \ref{app:binary}) is constant between successive eclipses of the same type, but not between primary and secondary eclipses for non-zero eccentricity.  For simplicity, we will allow $v_\text{x,bary}$ to be non-zero, but assume it to be constant in time (including between primary and secondary eclipses).

We will use Equation (\ref{eq:general_binary_vext}) and alternate the roles of the eclipsed and eclipsing stars, as necessary, using indices 1 and 2 to represent the primary and secondary stars, respectively.
\begin{equation}\begin{split}\label{eq:phase_separation_eccentric}
  \Delta \Phi_\text{sep, BATV} = \frac{1}{P c} \Biggl[ &
     \Delta z_{12}(t_\text{sec}) \left( \frac{1/q-1}{1/q+1} - \frac{1}{1/q+1} \frac{v_\text{x,bary}}{v_\text{x,1,orb}(t_\text{sec})} \right) \\ &-
     \Delta z_{21}(t_\text{pri}) \left( \frac{q-1}{q+1} - \frac{1}{q+1} \frac{v_\text{x,bary}}{v_\text{x,2,orb}(t_\text{pri})} \right)
  \Biggr] \>,
\end{split}\end{equation}
where $\Delta z_{12}$ and $v_{x,*,orb}$ can be found in terms of orbital elements in Appendix \ref{app:binary}.

\subsubsection{Circular Case}

We can make a few additional simplifications by examining the circular case.  Here, the separation between the two stars remains constant throughout the orbit, so $\Delta z_{12} (t_\text{sec}) = \Delta z_{21} (t_\text{pri}) = a \sin i$.  Additionally, the velocity of a given star is constant throughout the orbit, so the velocities in the x-direction are the same at primary and secondary eclipses, and therefore $v_\text{x,2,orb} \equiv v_\text{x,2,orb}(t_\text{pri}) = v_\text{x,2,orb}(t_\text{sec})$ and $v_\text{x,1,orb} \equiv v_\text{x,1,orb}(t_\text{pri}) = v_\text{x,1,orb}(t_\text{sec})$.  This then also allows us to use the mass-ratio to relate velocities via $v_\text{x,1,orb} = - q v_\text{x,2,orb}$:

\begin{equation}
  \Delta \Phi_\text{sep, BATV} = \frac{2 a \sin i}{P c} \left( \frac{1-q}{1+q} + \frac{1}{1+q} \frac{v_\text{x,bary}}{v_\text{x,2,orb}} \right) \>.
\end{equation}

We then represent $v_\text{x,2,orb}$ in terms of orbital elements (again, assuming the circular case, see Appendix \ref{app:binary} for $v_\text{x,2,orb}$ in the general, eccentric, case):

\begin{equation}
  v_\text{x,2,orb}(e=0) = \sqrt{ \frac{G M_\text{tot}} {a} } \frac{1}{1+q} \>.
\end{equation}

Then by using Kepler's third law, we can write the entire expression for the offset in phase-separation for the circular case in terms of $v_\text{x,bary}$, $q$, $i$, $M_\text{tot}$, and $a$:
\begin{equation}\label{eq:phase_separation_circular_a}
  \Delta \Phi_\text{sep, BATV} = \left( \frac{ G M_\text{tot} } { \pi^2 a } \right)^{1/2} \frac{\sin i}{c}
  \left[ \frac{1-q}{1+q} + v_\text{x,bary} \left( \frac{a}{G M_\text{tot}} \right)^{1/2} \right] \>,
\end{equation}
or in terms of $v_\text{x,bary}$, $q$, $i$, $M_\text{tot}$, and $P$:
\begin{equation}\label{eq:phase_separation_circular_p}
  \Delta \Phi_\text{sep, BATV} = \left( \frac{ 2 G M_\text{tot} } {\pi^2 P} \right)^{1/3} \frac{\sin i}{c}
  \left[ \frac{1-q}{1+q} + v_\text{x,bary} \left( \frac{P}{2 \pi G M_\text{tot}} \right)^{1/3} \right] \>.
\end{equation}

\begin{figure}[b!]
    \centering
    \includegraphics[width=0.5\textwidth]{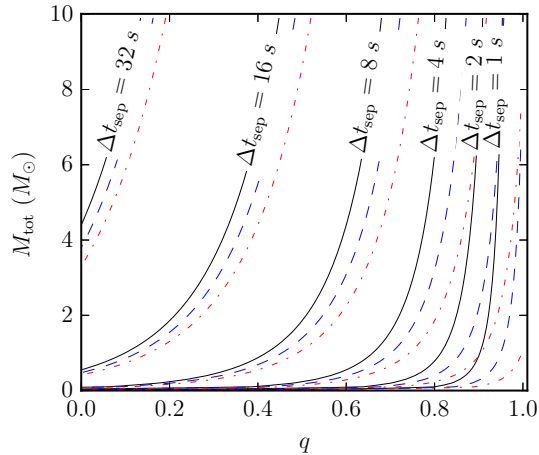}
    \caption{Change in eclipse separation (from the expected 0.5-phase) caused by BATV for a circular $i=90^\circ$ binary with a period of 1.0 d as a function of the total mass and mass-ratio shown in solid black for $v_\text{x,bary}/v_\text{x,f,orb,peri}= 0\%$, dashed blue for $5\%$, and dash-dotted red for $10\%$.}
    \label{fig:countour_separation}
\end{figure}

Figure \ref{fig:countour_separation} shows the magnitude of this shift from 0.5-phase separation for a circular binary with a period of 1.0 days converted to time units.  The magnitude increases as the mass-ratio becomes more extreme and as the total mass of the system increases.  For equal mass binaries with a period of 1 day, the shift caused by BATV is $\sim 1~s$.  However, for smaller mass ratios, the shift can reach $\sim 30~s$, which is easily observable with precision photometry.

\begin{figure}
    \centering
    \includegraphics[width=0.5\textwidth]{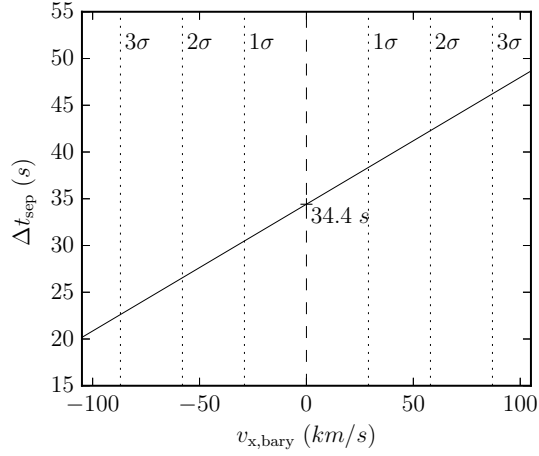}
    \caption{Change in transit separation (from the expected 0.5-phase, assuming a circular orbit), caused by BATV for KELT-9b as a function of $v_\text{x,bary}$.  With no barycentric transverse motion, the secondary event can still be expected to show a shift of $34.4~s$ due to its small mass ratio.  Black dotted vertical lines show the estimates for the distribution of $v_\text{x,bary}$ from GAIA for different confidence levels (see Figure \ref{fig:gaia_dist}).  At $3 \sigma$, 99.7 \% of objects will be influenced by the barycentric term of BATV by up to $\pm 12~s$.}
    \label{fig:separation_kelt9}
\end{figure}

For hot-Jupiters around fairly high-mass stars, for example, if an occultation can be observed and used to constrain the eccentricity, it is important to account for BATV in order to avoid misconstruing a phase-separation as non-zero eccentricity. It can also be important to account for a conservative uncertainty in the value of $v_\text{x,bary}$ and its influence on the phase-separation when determining measured uncerainties on the eccentricity or $e \cos \omega$.  KELT-9b \citep[Collins et al., in preparation]{Gaudi2017}, for instance, is a detected planet system which is particularly susceptible to BATV as it has a small mass-ratio of $q=0.0011$ and a fairly large total mass of $M_\text{tot} = 2.5~M_\odot$.  Although the system is not known to be exactly circular, it is expected to have been significantly circularized due to its short orbital period of $P=1.48~d$.  Figure \ref{fig:separation_kelt9} shows the expected time-shift of the secondary eclipse relative to the expected value as a function of $v_\text{x,bary}$.  For a reasonable range of transverse velocities adopted from the $3\sigma$ distribution from GAIA (see Figure \ref{fig:gaia_dist}), this shift could be anywhere from $\sim 20$ to $\sim 45$ seconds.  As the eccentricity of this system is not well-constrained, there will be a degeneracy in the contribution to this shift between BATV and a small, but non-zero, eccentricity.  With individual eclipses timed to a precision of $\sim 10~s$ (Collins et al, in preparation, private communication), the effect of BATV on the resulting uncertainties on $e \cos \omega$ could be to the same order as the effect of these timing uncertainties.

\citet{Kaplan2010} uses a similar expression, adapted from \citet{Fabrycky2010}, to constrain the mass-ratio of a double white dwarf binary.  He uses radial velocity data to determine the eccentricity of the orbit to be negligible and then uses the phase separation between primary and secondary eclipses along with the semi-amplitude of the single-lined radial velocities to determine the mass-ratio, $q$.  His Equation 4, reproduced below as Equation (\ref{eq:kaplan}) with the notation used in this work and divided by $P$ to translate from phase- to time-space, can be derived in the limit where $v_\text{x,bary} = 0$, and $i = 90^\circ$:  

\begin{equation}\label{eq:kaplan}
    \Delta \Phi_\text{sep, BATV}(v_\text{x,bary} \rightarrow 0, i \rightarrow 90^\circ) = \frac{1}{P} \left( \frac{ 2 G M_1 P^2}{\pi^2 c^3}\right)^{1/3} \frac{ \left( 1-q \right)}{\left(1+q\right)^{2/3}} \>.
\end{equation}

The above equations (\ref{eq:phase_separation_eccentric} for the general, eccentric, case and \ref{eq:phase_separation_circular_a} or \ref{eq:phase_separation_circular_p} when known to be circular) provide a more robust estimate of the phase-separation and therefore could be used in a similar matter to that of \citet{Kaplan2010} to provide constraints on the mass-ratio.  In practice, unless within a higher-order system, $v_\text{x,bary}$ will likely be unknown, in which case reasonable limits, or constraints adopted from GAIA proper motions, could be applied to estimate the resulting uncertainty on the mass ratio.

\subsection{Eccentric Systems with Apsidal Motion}\label{sec:scenarios_apsidal}

In the case of apsidal motion, the distance between the two components, $\Delta z_\text{bf}$, at a given eclipse (i.e.~primary or secondary) ranges throughout the entire precession cycle from $a (1-e) \sin i $ when the eclipse occurs at periastron to $a  (1+e) \sin i$ at apastron.  In addition, the velocity of the front star varies from $v_\text{f,orb,peri}$ at periastron to $v_\text{f,orb,peri} \left(1-e\right) / \left(1+e\right) $ at apastron.  As both the separation and velocities are time-dependent, the effect caused by the asymmetric velocities will also vary in time.  We can therefore determine the maximum peak-to-peak amplitude of this effect, $A_\text{BATV}$, over the whole apsidal motion cycle as the difference between Equation (\ref{eq:general_binary_vext}) expressed at periastron and apastron:

\begin{equation}\label{eq:apsidal}
    A_\text{BATV} = \left| \frac{2 e a \sin i}{c} \left(\frac{\xi-1}{\xi+1} - \frac{2}{\xi+1} \frac{v_\text{x,bary}}{v_\text{x,f,orb,peri}} \right) \right| \>.
\end{equation}

\begin{figure}
    \centering
    \includegraphics[width=0.5\textwidth]{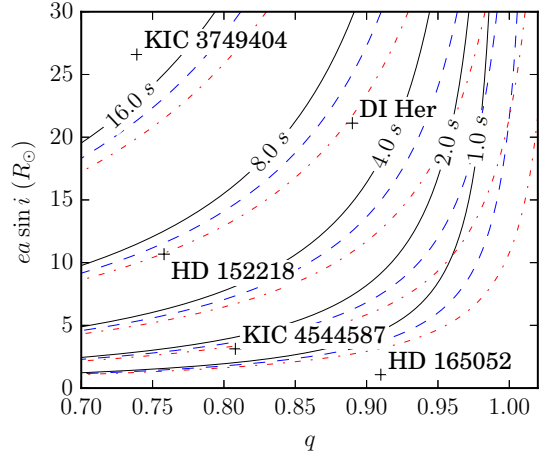}
    \caption{Peak-to-peak BATV amplitude (for primary eclipses, $\xi \equiv q$) over an entire apsidal motion cycle as a function of eccentricity ($e$) times projected semi-major axis ($a \sin i$) and mass ratio ($q$) shown in solid black for $v_\text{x,bary}/v_\text{x,f,orb,peri}=0\%$, dashed blue for $1\%$, and dot-dashed red for $2\%$.  Also included are several known apsidal motion systems, whose adopted values and citations are provided in Table \ref{table:apsidal_binaries}.}
    \label{fig:contour_apsidal_binaries}
\end{figure}

\begin{figure}
    \centering
    \includegraphics[width=0.5\textwidth]{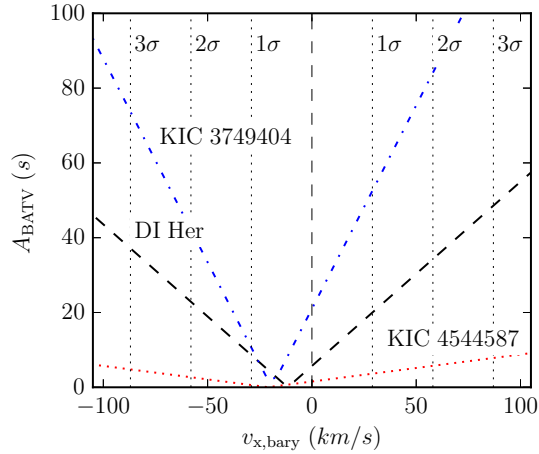}
    \caption{Peak-to-peak BATV amplitude (for primary eclipses, $\xi \equiv q$) over an entire apsidal motion cycle for several known apsidal motion systems as a function of the bulk transverse velocity, $v_\text{x,bary}$.  DI Her is shown in dashed black, KIC 3749404 in dot-dashed blue, and KIC 4544587 in dotted red.  Black dotted vertical lines show the estimates for the distribution of $v_\text{x,bary}$ from GAIA for different confidence levels (see Figure \ref{fig:gaia_dist}).}
    \label{fig:apsidal_vs_vext}
\end{figure}

\begin{figure}
    \centering
    \includegraphics[width=0.5\textwidth]{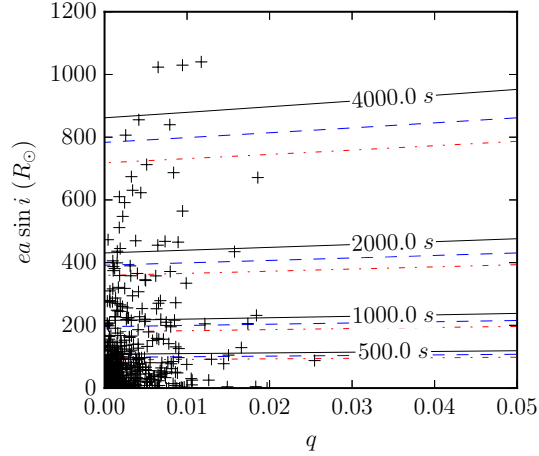}
    \caption{Same as Figure \ref{fig:contour_apsidal_binaries} but extended to low mass-ratios for planets.  Shown are the peak-to-peak BATV amplitudes (for transits) over an entire apsidal motion cycle as a function of eccentricity ($e$) times projected semi-major axis ($a \sin i$) and mass ratio ($q$) for all confirmed Kepler exoplanets.   The contours are shown in solid black for $v_\text{x,bary}/v_\text{x,f,orb,peri}=0\%$, dashed blue for $5\%$, and dot-dashed red for $10\%$.  Note that these are not necessarily known to exhibit apsidal motion, but do represent the parameter space of known exoplanets.}
    \label{fig:contour_apsidal_exoplanets}
\end{figure}

Equation (\ref{eq:apsidal}) is plotted in Figure \ref{fig:contour_apsidal_binaries} for several values of $v_\text{x,bary}/v_\text{x,f,orb,peri}$ along with several known apsidal motion cases, whose adopted parameters are listed in Table \ref{table:apsidal_binaries}.  The parameter space in which this effect is maximized (i.e.~small mass-ratio, large eccentricity, large semi-major axis) also minimizes the chance of observing and detecting the eclipses.  Largely because of this, most known apsidal motion binaries have fairly small contributions when assuming no barycentric transverse velocity.  However, it is not implausible to imagine a system being observed in which it is necessary to account for BATV in order to accurately determine the true precession rate,  particularly as missions such as GAIA begin to give us constraints on the barycentric transverse velocities of these systems.  This will require a fairly long baseline, as all cases listed in Table \ref{table:apsidal_binaries} have contributions from BATV below 0.1 seconds per day (even with conservative estimates on $v_\text{x,bary}$, see Figure \ref{fig:apsidal_vs_vext}).

Figure \ref{fig:apsidal_vs_vext} shows this same amplitude for these apsidal motion binaries as a function of $v_\text{x,bary}$ by computing $v_\text{x,f,orb,peri}$ for each binary using the total mass, $M_\text{tot}$, adopted from the literature as listed in Table \ref{table:apsidal_binaries}.  Note that even for no barycentric transverse velocity, the finite speed of light still requires a corrective term to apsidal motion for non equal-mass systems.  Barycentric transverse motion does, however, contribute significantly even at relatively low velocities.  Also note that since $\xi < 1$ for all of these cases (see footnote in Table \ref{table:apsidal_binaries}), the two terms in Equation (\ref{eq:apsidal}) are opposite in sign for small negative barycentric velocities, therefore decreasing the amplitude of the effect until the second term eventually dominates (see Figure \ref{fig:apsidal_vs_vext}).  Once these transverse velocities are known, BATV may then become a significant contribution for some systems.

\begin{deluxetable}{rrrrrrrl}
\tablewidth{\textwidth}
\tablecaption{Adopted values and computed amplitudes for known apsidal motion binaries\label{table:apsidal_binaries}}
\tablehead{
   \colhead{System} & \colhead{e} & \colhead{a sini} & \colhead{q} & \colhead{$M_\text{tot}$} & \colhead{$P_\text{apsidal}$} & \colhead{$A_\text{BATV}$} & \colhead{Reference} \\
   \colhead{} & \colhead{} & \colhead{($R_\odot$)} & \colhead{} & \colhead{($M_\odot$)} & \colhead{($d$)} & \colhead{($s$)} & \colhead{} \\
}

\startdata
DI Herculis & 0.489 & 43.2 & 0.89 & 9.7 & 55400 & 5.7 & \citet{guinan1985} \\
HD 152218\tablenotemark{$\ast$} & 0.269 & 39.7 & 0.76\tablenotemark{$\dagger$} & \nodata\tablenotemark{$\ddagger$} & 176 & 6.8 & \citet{rauw2016} \\
HD 165052\tablenotemark{$\ast$} & 0.090 & 11.9 & 0.91 & \nodata\tablenotemark{$\ddagger$} & 30 & 0.2 & \citet{ferrero2013} \\
KIC 3749404 & 0.659 & 40.4 & 0.74 & 3.1 & 309 & 18.5 & \citet{hambleton2016} \\
KIC 4544587 & 0.288 & 10.8 & 0.81 & 3.6 & 182 & 1.5 & \citet{hambleton2013} \\
\enddata

\tablenotetext{\ast}{spectroscopic binary - may not eclipse.}
\tablenotetext{\dagger}{reported as $q=1.32$ in \citet{rauw2016}.}
\tablenotetext{\ddagger}{not included as only $M_\text{tot} \sin^3 i$ is known.}

\tablecomments{All values except BATV amplitudes are either directly or computed from values in the cited reference.  All reported transverse amplitudes are computed for $v_\text{x,bary}=0$.  See Figures \ref{fig:contour_apsidal_binaries} and \ref{fig:apsidal_vs_vext} to see the dependence of these values on the barycentric transverse velocity.}

\end{deluxetable}

In the case of exoplanets, the mass-ratio will be small, resulting in a large contribution even when the size of the orbit is small.  Figure \ref{fig:contour_apsidal_exoplanets} shows the same as Figure \ref{fig:contour_apsidal_binaries}, but for the parameter space of known Kepler exoplanets.  Note that these are not necessarily known apsidal motion cases, but the figure does exhibit that BATV can be quite significant for any exoplanet exhibiting precession.

\subsection{Binary System with Change in Inclination}\label{sec:scenarios_inclchange}

There are several known cases in which the inclination of an eclipsing system changes quickly enough to cause an observable change in the depth of the eclipse, including AY Mus \citep{Soderhjelm1974}, V907 Sco \citep{Lacy1999}, SS Lac \citep{Torres2001}, and a number of systems in the Magellanic Clouds \citep{Jurysek2017}.  In some of these cases, this change in inclination is so extreme that eclipse can be seen to begin or cease entirely.  A change in inclination can be due to any external forces on the system, including the presence of any additional bodies in the system causing dynamical effects, including Kozai cycles \citep{Kozai1962, Mazeh1979}.  Note that these dynamical effects may also cause perturbations to other orbital elements which could result in additional contributions to the shape and timing of eclipses.

Similar to the apsidal motion case, a change in inclination also results in a change in the projected separation of the two stars between successive eclipses, but the velocities at eclipse remain fixed.  In this case, the separation at eclipse will vary from $s_\text{ecl} \equiv \Delta z_\text{bf}(i=90^\circ)$ to $\Delta z_\text{bf} (i=0^\circ) = 0$.  However, as eclipse times are only measurable when eclipses are still present, the maximum observed amplitude will only occur between some critical inclination, $i_\text{crit}$, and $90^\circ$.

This critical inclination can be approximated geometrically as follows:
\begin{equation}
    i_\text{crit} = \cos^{-1} \left( \frac{R_\text{f} + R_\text{b}}{s_\text{ecl}} \right) \>,
\end{equation}
where $R_\text{f}$ and $R_\text{b}$ are the radii of the front and back stars, respectively, and $s_\text{ecl}$ is the (non-projected) distance between the two components at eclipse (i.e.~$a$ for circular binaries).  Therefore the observable effect can be approximated by:


\begin{equation}\label{eq:inclination}
    A_\text{BATV} = \left| \left[ 1 - \sqrt{1 - \left(\frac{R_\text{f} + R_\text{b}}{s_\text{ecl}} \right)^2 } \right] \frac{s_\text{ecl}}{c} \left( \frac{\xi-1}{\xi+1} - \frac{1}{\xi + 1 } \frac{v_\text{x,bary}}{v_\text{x,f,orb}} \right) \right| \>.
\end{equation}

\begin{figure}
    \centering
    \includegraphics[width=0.5\textwidth]{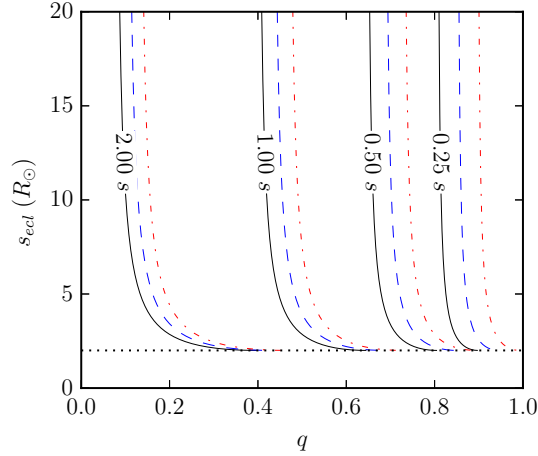}
    \caption{BATV amplitude (for primary eclipses, $\xi \equiv q$) over a change in inclination from $i_\text{crit}$ to $90^\circ$ as a function of the mass-ratio ($q$) and the separation between two components at eclipse ($s_\text{ecl}$) for a system in which the sum of radii is $R_\text{f} + R_\text{b} = 2 R_\odot$.  The contours are in solid black for $v_\text{x,bary}/v_\text{x,orb}=0\%$, dashed blue for $5\%$, and dot-dashed red for $10\%$.  The dashed black line at $s_\text{ecl} = 2 R_\odot$ represents the limit at which the two stars will be in contact.}
    \label{fig:contour_inclination}
\end{figure}

The expression above is plotted in Figure \ref{fig:contour_inclination} for the case where  $R_\text{f} + R_\text{b} = 2 R_\odot$, showing that, for a binary with a change in inclination, BATV can have a measurable contribution to the ETVs on the order of seconds, with any barycentric transverse velocity potentially increasing the magnitude of the effect. As was the case for apsidal motion, $\xi < 1$ will result in opposing signs for the two terms on the right in Equation (\ref{eq:inclination}), and therefore a small negative $v_\text{x,bary}$ will actually decrease the overall amplitude before eventually dominating.

\subsection{Hierarchical Triple Systems}\label{sec:scenarios_triple}

For a hierarchical triple system in which a third star is in orbit with an inner-binary system, the barycentric transverse velocity of the inner-binary system varies in time throughout the period of the outer-orbit, resulting in a cyclical contribution to the ETVs of the inner-binary caused by BATV.

Here $v_\text{x,bary}$ in Equation (\ref{eq:general_binary_vext}) becomes the transverse velocity of the barycenter of the inner-binary caused by its orbit about the barycenter of the entire triple system and $v_\text{x,f,orb}$ is the velocity of the eclipsing star caused by the inner-orbit alone, projected along the instantaneous direction of $v_\text{x,bary}$ (by definition of the x-direction).  As we know the barycenter of the inner-binary is moving in the z-direction throughout the outer-orbit, the assumptions in Equation (\ref{eq:general_binary_vext}) can no longer be assumed.  Note though that the z-velocity of the inner-binary is minimized as the contribution from BATV is maximized, and vice versa.  Nevertheless, these assumptions should be dropped and a numerical method or the linear-approximation explained in Appendix \ref{app:binary_linear} should be used to determine precise times of observed eclipse as a function of time.  These equations, along with the orbital elements provided in Appendix \ref{app:triple}, can still be particularly useful in conjunction with classical LTTE and dynamical equations to fit orbital elements of the outer-orbit to observed ETVs of an inner eclipsing binary prior to completing a full dynamical model with light time delay.

\subsubsection{Circular Coplanar Case}

For simplicity, to compare the contribution to the ETVs of BATV to both LTTE and dynamical effects, we'll examine the case of a hierarchical triple system in which both orbits are circular and share the same plane (i.e.~$i_\text{in} = i_\text{out}$ and $\Omega_\text{in} = \Omega_\text{out}$).   Coplanar orbits maximize the contribution of BATV as the barycentric transverse velocity caused by the motion around the center-of-mass of the entire system is most aligned with the velocity of the stars in the inner-binary at eclipse.  We derive the circular case using the following conditions:
\begin{equation}
\begin{split}
    \Delta z_\text{bf} (e_\text{in}=0) &= a_\text{in} \sin i_\text{in} \>, \\
    v_\text{x,f,orb} (e_\text{in}=0, eclipse) &= \sqrt{ \frac{G M_{12} } {a_\text{in}} } \frac{1}{1+\xi_\text{in}} \>, \text{and}\\
    v_\text{x,bary} (e_\text{out}=0) &=  \cos (\upsilon_\text{12}) \sqrt{ \frac{G M_{123} } {a_\text{out}} }  \frac{q_\text{out}}{1+q_\text{out}}\>,
\end{split}
\end{equation}
where the subscript ``in'' represents the inner-orbit and ``out'' the outer-orbit in which the inner-binary is the primary component ($q_\text{out} \equiv M_{3} / M_{12}$) and is treated as a point mass at its own barycenter.  $\upsilon_\text{12}$ is then the true anomaly of the inner binary within the outer-orbit.

Substituting these into Equation (\ref{eq:general_binary_vext}), we can get the contribution of BATV throughout the outer-orbit as a function of the mass-ratios and semi-major axes:
\begin{equation}\label{eq:triple_circular_coplanar}
    \Delta t_\text{BATV} (e=0, \text{coplanar}) =  \frac{a_\text{in} \sin i_\text{in}}{c} \left[ \frac{\xi_\text{in} - 1}{\xi_\text{in} + 1} \pm \cos (\upsilon_\text{12}) \left(\frac{a_\text{in}}{a_\text{out}}\right)^{1/2}  \frac{q_\text{out}}{(1+q_\text{out})^{1/2}} \right] \>,
\end{equation}
where the sign on the second term is positive for a prograde orbit and negative for a retrograde orbit.

The cosine term above varies in sign as $v_\text{x,bary}$ flips direction throughout the outer-orbit.  The peak-to-peak amplitude is therefore the difference between this expression taken while the inner binary is in the front and back of the outer-orbit, i.e.~$\Delta t_\text{BATV} (\upsilon_\text{12} = 0) - \Delta t_\text{BATV} (\upsilon_\text{12} = \pi)$.  Since the extrema used in the amplitude are taken at points along the outer-orbit in which the inner-binary is not moving in the z-direction, this amplitude can safely be determined without the need for numerical computations.  As only the second term is time-dependent, the peak-to-peak amplitude does not depend on $\xi_\text{in}$:

\begin{equation}\label{eq:triple_amp}
    A_\text{BATV} (e=0, \text{coplanar}) = \pm 2 \frac{a_\text{in} \sin i_\text{in}}{c} \left( \frac{a_\text{in}}{a_\text{out}} \right)^{1/2}  \frac{q_\text{out}}{(1+q_\text{out})^{1/2}} \>.
\end{equation}

\subsubsection{ETV Contribution Compared to LTTE}

Classical LTTE for the same circular, coplanar, case will contribute peak-to-peak ETVs equivalent to the photon travel time across the outer orbit:
\begin{equation}
    A_\text{LTTE} (e=0) = 2 \frac{a_\text{out} \sin i_\text{out}}{c} \frac{q_\text{out}}{1+q_\text{out}} \>.  
\end{equation}
Since we are exploring the coplanar case, we can set $i_\text{out} = i_\text{in}$, and can therefore approximate the ratio between the BATV and LTTE contributions to the ETVs as follows:

\begin{equation}\label{eq:triple_perc}
    \frac{A_\text{BATV} (e=0, \text{coplanar})}{A_\text{LTTE} (e = 0)} =  \pm \left( \frac{a_\text{in}}{a_\text{out}} \right)^{3/2} \left( 1 + q_\text{out} \right)^{1/2}  \>.
\end{equation}

\begin{figure}
    \centering
    \includegraphics[width=\textwidth]{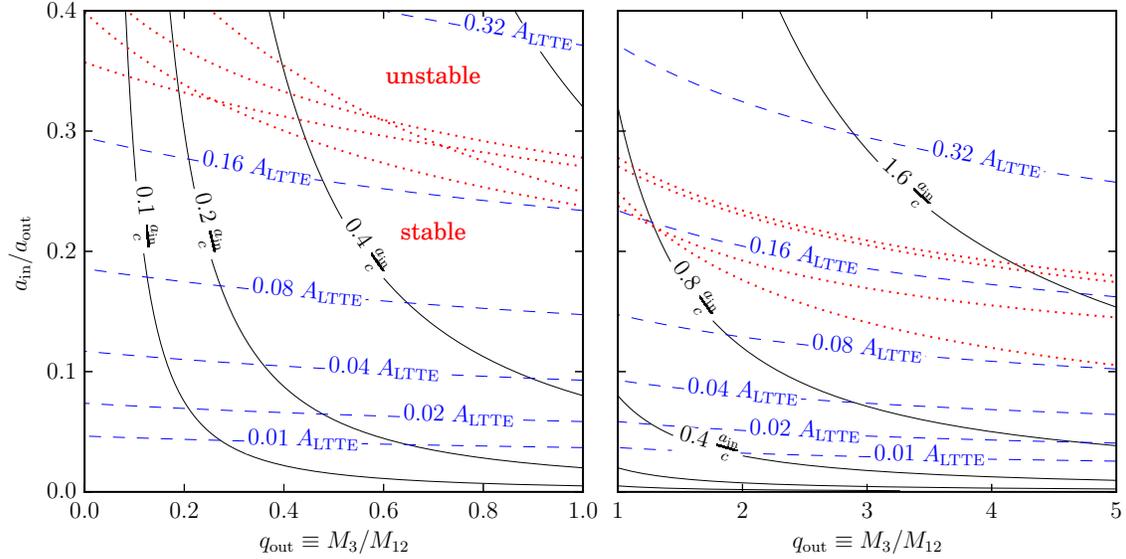}
    \caption{Peak-to-peak amplitude of BATV over an entire orbit of the inner-binary within the outer-binary for the circular and coplanar case, viewed edge-on at $i=90^\circ$. The left shows $q_\text{out}<1$ while the right shows $1<q_\text{out}<5$.  The solid black contours are in terms of the photon travel time between the two eclipsing components in the inner-binary, $a_\text{in}/c$, from Equation (\ref{eq:triple_amp}) and the dashed blue contours are in terms of the ratio of the amplitude as compared to classical LTTE, $A_\text{LTTE}$, from Equation (\ref{eq:triple_perc}).  The red dotted lines represent the estimated stability limits, for $q_\text{in}=1$, according to \citet{harrington1972}, \citet{bailyn1987}, \citet{eggleton1995}, and \citet{mardling2001} as compiled by \citet{mikkola2008}.}
    \label{fig:contour_triples_ltte}
\end{figure}

Figure \ref{fig:contour_triples_ltte} shows the magnitude of BATV for circular, coplanar, hierarchical orbits.  This effect is maximized as $q_\text{out} \rightarrow \infty$ (so that the inner-binary's velocity through space is increased) and as $a_\text{in} \rightarrow a_\text{out}$ (the more tightly packed the system is, the larger the ratio between barycentric and orbital transverse velocity for the inner-binary).  Also depicted in Figure \ref{fig:contour_triples_ltte} are various estimates for the stability limit of hierarchical triple systems according to \citet{harrington1972}, \citet{bailyn1987}, \citet{eggleton1995}, and \citet{mardling2001} as compiled and summarized by \citet{mikkola2008}.  Generally speaking, in the most extreme but still stable scenarios, it is possible for BATV to contribute $\approx 15-20 \%$ that of classical LTTE.  In the most stable hierarchical systems, however, it is likely that the contribution from BATV will be under $1 \%$ that of LTTE.

\begin{figure}
    \centering
    \includegraphics[width=\textwidth]{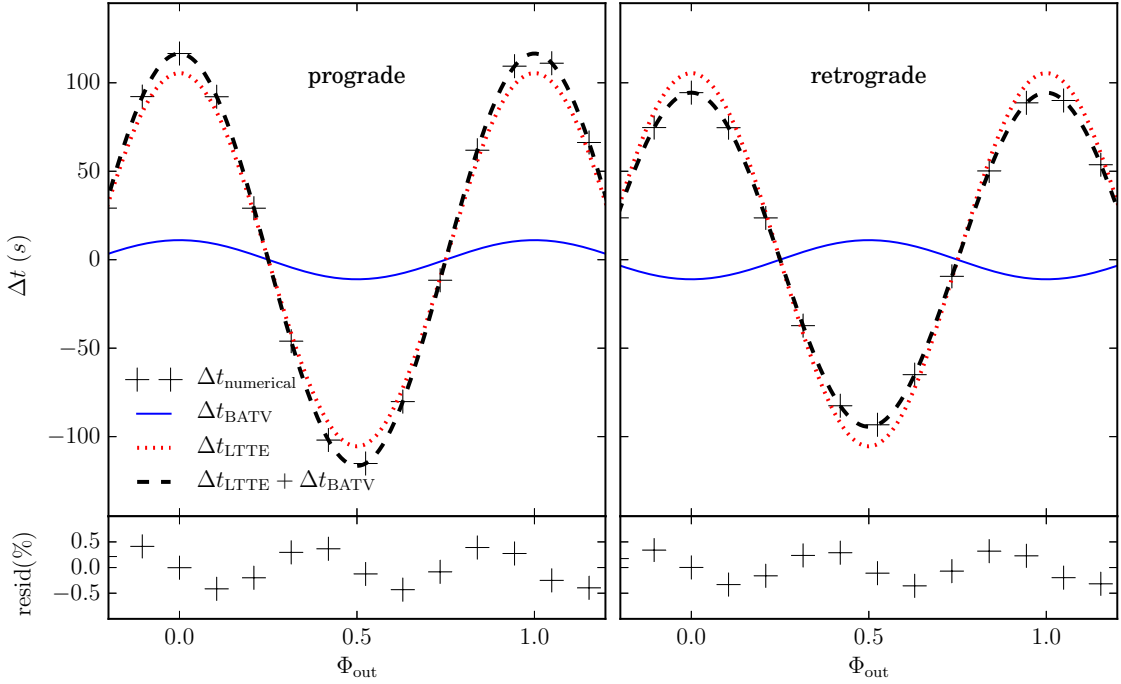}
    \caption{Contribution to ETVs by both classical LTTE and the BATV effect for a circular coplanar hierarchical triple, for the case where $q_\text{out}=10$ and $a_\text{in}/a_\text{out} = 0.1$.  On the left is the prograde case, showing an increase in the overall magnitude of the ETVs, whereas the right shows the retrograde case with a decrease in the magnitude.  The expressions are plotted here as a continuous function of $\Phi_\text{out}$ but note that they are only applicable at times at which an eclipse of the inner-binary occurs, shown as black +s for the numerical solution.  The residuals between the exact numerical solution and the analytic expression (due to the stated approximations) are shown in the lower panel, reaching up to 0.5\% of $A_\text{BATV}$.}
    \label{fig:etv_contribution}
\end{figure}

Nevertheless, without properly accounting for BATV, fitting the LTTE contribution of ETV observations would result in an incorrect measurement of the amplitude of the timing variations caused by LTTE.  Since $A_\text{LTTE} \propto  P_\text{out}^{2/3} \left( m_3 / m_{123}^{2/3} \right)$, this will result in an overestimate or underestimate in the mass-ratio (and therefore mass of the third body) for prograde and retrograde orbits, respectively (see Figure \ref{fig:etv_contribution}).

Figure \ref{fig:etv_contribution} also compares the analytical approximation for LTTE and BATV in Equation \ref{eq:triple_circular_coplanar} (assuming nested Keplerian orbits) to the exact numerical solution.  The residuals in the case shown are on the order of $1 \%$ the amplitude of the BATV contribution and are caused by the approximations used: that the barycenter of the inner-binary does not move in the z-direction and that the eclipsing stars travel in constant and straight trajectories during the photon travel time.  When the systematic residuals due to these approximations prove too significant to neglect, Equations (\ref{eq:condition_eclipse}-\ref{eq:condition_t0}) can be solved iteratively in conjunction with the relevant equations of motion.  Provided that $\vec{r}_b(t)$, $\vec{r}_f(t)$, and $\vec{r}_\mathrm{bc}(t)$ can be computed, the scheme is as follows:
\begin{itemize}
    \item pick a timestamp $t_\text{b}$ (e.g.~$t_\text{align}$) and compute $\vec{r}_\text{b}(t_\text{b})$;
    \item solve Equation (\ref{eq:condition_eclipse}) for $t_f$; in most cases this needs to be done iteratively, i.e.~by employing a Newton-Raphson method;
    \item given $t_\text{f}$, compute $\vec{r}_\text{f}(t_\text{f})$;
    \item given $\vec{r}_\text{b}(t_\text{b})$ and $\vec{r}_\text{f}(t_\text{f})$, calculate the difference between both sides of Equation (\ref{eq:condition_ltte}), $\Delta = t_\text{f} - t_\text{b} - [z_\text{f}(t_\text{f}) - z_\text{b}(t_\text{b})]/c$;
    \item iterate the scheme over $t_\text{b}$ until $\Delta \to 0$ to a required level of precision;
    \item given $t_\text{f}$ and $\vec{r}_\text{bc} (t_0)$, solve iteratively for $t_0$ using Equation (\ref{eq:condition_t0}).
\end{itemize}

\subsubsection{ETV Contribution Compared to Dynamical Effects}

Similarly to BATV, dynamical effects increase as the triple system becomes more tightly packed, and therefore also maximize their contribution to the ETVs.  The amplitude of this effect can be approximated \citep[see][]{mayer1990, Borkovits2003, Borkovits2011, Rappaport2013} as:
\begin{equation}
\begin{split}
    A_\text{dyn} &= \frac{3}{8 \pi} \frac{M_3}{M_{123}} \frac{P_\text{in}^2}{P_\text{out}} (1-e^2)^{-3/2} \\
    &= \frac{3}{4} G^{-1/2} q_\text{out} a_\text{in}^3 a_\text{out}^{-3/2} M_{123}^{-1/2} (1-e^2)^{-3/2} \>.
\end{split}
\end{equation}

We can then determine the ratio between BATV and dynamical contributions to the ETVs for the circular, coplanar, edge-on case:
\begin{equation}\label{eq:triple_dyn_perc}
    \frac{A_\text{BATV} (e=0, i=90^\circ, \text{coplanar})}{A_\text{dyn} (e = 0)} = \frac{8}{3} \frac{G^{1/2}}{c} a_\text{in}^{-3/2} a_\text{out} M_\text{12}^{1/2}   \>.
\end{equation}

\begin{figure}
    \centering
    \includegraphics[width=0.5\textwidth]{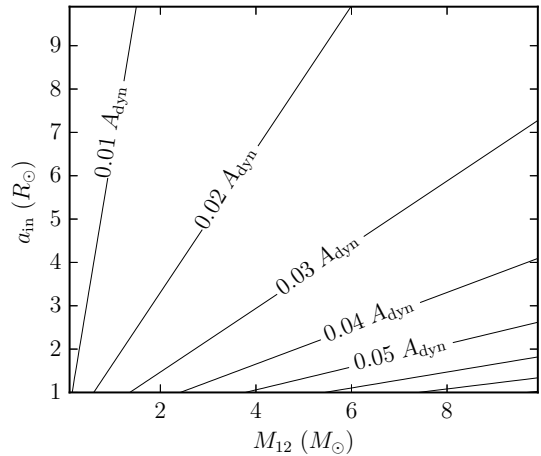}
    \caption{Contribution of the peak-to-peak amplitude of BATV as compared to $A_\text{dyn}$ for the circular coplanar case in which $a_\text{out}$ is fixed to be at the mean stability limit (see text for more details).  Increasing $a_\text{out}$ (i.e.~increasing the stability of the system) will increase this relative contribution linearly.}
    \label{fig:contour_triples_dyn}
\end{figure}

Note that, unlike for LTTE, this ratio does not depend on $q_{out}$ but instead on the total mass of the inner binary, $M_\text{12}$.  Figure \ref{fig:contour_triples_dyn} shows this ratio with $a_\text{out}$ at the mean stability limit from \citet{harrington1972}, \citet{bailyn1987}, \citet{eggleton1995}, and \citet{mardling2001}, assuming $e_\text{in}=0$, $e_\text{out}=0$, and $q_\text{in}=1$ (Figure \ref{fig:contour_triples_ltte} shows the discrepancy between these models and the relation with $q_\text{in}$).  As this ratio (Equation \ref{eq:triple_dyn_perc}) scales linearly with $a_\text{out}$, the contribution from BATV relative to dynamical effects will increase for increasingly stable systems.

For any system, it is likely that either LTTE or dynamical effects will dominate over BATV (see Figure 7 in \citet{Rappaport2013} for a comparison between $A_\text{LTTE}$ and $A_\text{dyn}$ as a function of $P_\text{out}$).  However, it may still be necessary to account for BATV in order to achieve accurate and precise determinations on the system parameters.

\section{Conclusion}\label{sec:conclusion}

Barycentric and asymmetric transverse velocities (BATV) influences the observed timing of eclipses, with respect to the barycentric frame of reference.  As mentioned previously, this should be applied in addition to any necessary barycentric light time (i.e.~LTTE) and dynamical corrections.  For the purposes of fitting an approximate analytical equation to ETVs or TTVs, the validity of the assumptions in the equations above should be considered.  For complete accuracy, Equations  (\ref{eq:condition_eclipse}), (\ref{eq:condition_ltte}), and (\ref{eq:condition_t0}) should be solved directly using the known equations of motion or a numerical integrator, if possible.

Even without external barycentric motion, any internal asymmetry in the transverse velocities during eclipse (caused by a non-unity mass ratio or non-zero eccentricity) can still introduce a shift in the observed eclipse times.  If any of these terms could be time-dependent, it is important not to neglect the contribution of BATV towards the overall observed timing of an eclipse.  In any cases where the time-dependence of these values or the barycentric transverse velocity is unknown, this effect should, at the very least, be folded into resulting uncertainties by assuming conservative upper limits on all time derivatives and velocities.

The prospect of measuring the plane-of-sky barycentric velocity for a given system is improving as GAIA continues to release parallaxes and proper motions \citep{gaia2016a, gaia2016b, Lindegren2016}.  In order to make use of these proper motions, the orientation of the binary on the sky (relative to the proper motion) must be constrained.  In most cases this is not tractable, but proper motions can still be used to provide an upper limit on BATV. \citet{Ofir2014} notes that the projected position angle of an orbit on the sky can be determined by comparing simultaneous eclipse times from widely separated observers (i.e.~from earth and a space telescope) along with high-precision parallax.  In conjunction with GAIA proper motions, the external transverse barycentric velocity may be attainable.  Whenever possible, a known value for the external transverse barycentric motion can greatly help in determining the true \emph{physical} cause of any ETV signal.

In addition to eclipse timings, an asymmetry in velocities can also be expected to influence eclipse durations as well as the shape of the overall eclipse profile.  Any photodynamical light curve modeling code that accounts for the finite speed of light, should account for these effects automatically.  Specifically for the case of triple stellar systems, Conroy et al., in prep will discuss the implementation of light time effects and eclipse timing variations within PHOEBE \citep{Prsa2016}.

\acknowledgements

K.~E.~Conroy is supported under NASA NESSF Fellowship \#NNX15AR87H.  We also acknowledge support from the NSF AAG grant \#1517474. The authors would like to thank K.~M.~Hambleton, A.~Kochoska, M.~Lund, K.~Collins, Eric Agol, and Scott Gaudi for very helpful discussions and comments. This research has made use of the NASA Exoplanet Archive, which is operated by the California Institute of Technology, under contract with the National Aeronautics and Space Administration under the Exoplanet Exploration Program.  This work has made use of data from the European Space Agency (ESA) mission {\it Gaia} (\url{https://www.cosmos.esa.int/gaia}), processed by the {\it Gaia} Data Processing and Analysis Consortium (DPAC, \url{https://www.cosmos.esa.int/web/gaia/dpac/consortium}). Funding for the DPAC has been provided by national institutions, in particular the institutions participating in the {\it Gaia} Multilateral Agreement.

\appendix

\section{Estimate of Error Introduced by Constant Velocity Approximation}\label{app:approximation}

We can estimate how strongly the assumption of constant transverse velocities throughout the light travel time between the two eclipsing components affects the resulting timings by determining how much the velocities change from the time of photon emission until interception. Let us quantify the change in the direction of star's velocity $\hat {\bf v}$ by its projection onto the direction velocity at geometric alignment $\hat {\bf v}_{\rm align}$:
\begin{equation}
  |\hat {\bf v}\cdot \hat {\bf v}_{\rm align}| = \cos(\alpha_{\rm ch}) \>,  
\end{equation}
with $\alpha_{\rm ch}$, measuring the change in angle between the vectors. In an isolated binary system this angle for any of the stars involved is at most of the order of magnitude of:
\begin{equation}
  \alpha_{\rm ch,max} \approx \frac{a \omega}{c \sqrt{1-e^2} (1 + \min\{\xi,1/\xi\})}\>,
\end{equation}
with $\omega$ being the orbital angular frequency, $a$ the semi-major axis and $e$ the eccentricity of the orbit. This means that a faster orbital velocity and a larger separation will generally yield a larger change in velocity throughout the time interval for either star, resulting in a larger error introduced by making this assumption.

\section{Linear Motion Along Line-of-Sight}\label{app:binary_linear}

Let us assume that both stars move linearly near the time of mid-eclipse in the $xz$ plane, with their trajectory expressed as:
\begin{equation}
x_i (t) = v_{\text{x},i} t\>, \qquad
z_i (t) = z_i(0) + v_{\text{z},i} t \qquad{\rm for}\qquad i = {\rm b},{\rm f}\>,
\label{eq:lin_motion}
\end{equation}
and the barycentric reference plane, $z_0$, moves along the $z$-axis as:

\begin{equation}
    z_0(t) = z_0(0) + v_{z,0} t \>.
\end{equation}

The equations of linear motion (\ref{eq:lin_motion}) along with Equations (\ref{eq:condition_ltte}) and (\ref{eq:condition_t0}) represent a system of linear equations which can be solved for $t_0$:
\begin{equation}\label{eq:general_linear_z}
    t_0 = 
        \left[1 - \frac{v_{z,0}}{c} \right]^{-1} 
        \left[ t_{\rm f} \left (1  - \frac{v_{z,{\rm f}}}{c} \right) - \frac{z_{\rm f}(0) - z_0(0)}{c} \right] \>.
\end{equation}
The barycenter's motion is given by:
\begin{equation}
    z_0(t) = \frac{z_\text{b}(t) + z_\text{f}(t) \xi} {1 + \xi} \>,
\end{equation}
where $\xi \equiv M_{\rm f}/M_{\rm b}$ is the mass ratio, $q$, for a primary eclipse or the inverse of the mass ratio, $1/q$ for a secondary eclipse. The time shift, relative to the time of alignment, is then equal to:
\begin{equation}\label{eq:binary_linear_z}
    \begin{split}
    \Delta t_\text{BATV}
    = 
    \frac{ z_{\rm f} (0) - z_{\rm b}(0)}{c}
    & \left[ 
    1 - \frac{v_{z,{\rm b}} + \xi v_{z,{\rm f}}}{(1+\xi)c}
    \right]^{-1} \\
    & \left[
    \frac
        {v_{x,{\rm f}} c} 
        {c(v_{x,{\rm b}} - v_{x,{\rm f}}) + v_{x,{\rm f}} v_{z,{\rm b}} - v_{x,{\rm b}}v_{z,{\rm f}}} 
    \left (
        1 - \frac{v_{z,{\rm f}}}{c} 
    \right)
    - \frac{1}{1+\xi}
    \right]\>.
    \end{split}
\end{equation}
In the typical situation in which all velocities are much smaller than speed of light, $|v_{z,i}| \ll c$, we may approximate Equation (\ref{eq:binary_linear_z}) as:
\begin{equation}
    \begin{split}
    \Delta t_\text{BATV} \approx  \frac{z_{\rm f} (0) - z_{\rm b}(0)}{c} 
    &\bigg [
    \frac{\xi v_{x,{\rm b}} + v_{x,{\rm f}} }{(1+\xi) (v_{x,{\rm f}} - v_{x,{\rm b}})} + \\
    &\frac{
        (v_{x,{\rm b}} - v_{x,{\rm f}})
        (\xi v_{x,{\rm b}} + v_{x,{\rm f}})
        (v_{z,{\rm b}} + \xi v_{z,{\rm f}})  
        -(1+\xi)^2 v_{x,{\rm b}} v_{x,{\rm f}} (v_{z,{\rm b}} - v_{z,{\rm f}})
    }
    {c(1+\xi)^2 (v_{x,{\rm f}} - v_{x,{\rm b}})^2} +\\
    & O(c^{-2})\bigg ]\>.
    \end{split}
\end{equation}
Note that the leading order approximation is identical to Equation (\ref{eq:general_constant_z}).

\section{Orbital Elements for Binary Systems}\label{app:binary}

We can write several of the terms in Equation (\ref{eq:general_binary_vext}) with orbital elements of a binary star system.

For the velocity of the front star (f) in our binary (fb), we want the velocity projected along the longitude of ascending node.  That can be represented as follows:
\begin{equation}\label{eq:orbital_binary_vx}
\begin{split}
    v_\text{x,f,orb} = \sqrt{\frac{G}{M_\text{fb} a_\text{fb} (1-e_\text{fb}^2)}} M_\text{b} \Biggl[ & -\sin (\nu_{f} (t)) \cos \omega_\text{fb} \\ &+ \left[ \cos (\nu_{f} (t)) +  e_\text{fb} \right] \sin \omega_\text{fb} \Biggr] \>,
\end{split}
\end{equation}
where $\nu (t)$ is the true anomaly of a given star at time $t$, $M$ is mass, $\omega$ is the argument of periastron, $a$ is the semi-major, and $e$ is the eccentricity of the orbit.

The separation between the front and back components projected along the line-of-sight, $\Delta z_\text{fb}$, can be written as follows:

\begin{equation}\label{eq:orbital_binary_sep}
    \Delta z_\text{fb} = \frac{a_\text{fb} (1-e_\text{fb}^2)}{1+e_\text{fb} \cos (\nu_\text{f} (t))} \sin _\text{fb} \left[ \cos (\omega_\text{fb} + \nu_\text{f}(t)) + e_\text{fb} \cos \omega_\text{fb} \right] \>.
\end{equation}

\section{Orbital Elements for Hierarchical Triple Systems}\label{app:triple}

In the case of a hierarchical triple system, Equation (\ref{eq:general_binary_vext}) can be expressed in terms of orbital elements of both the inner and outer Keplerian orbits.

For the velocity of the eclipsing component (f) in our inner-binary (fb), we want the velocity projected along the longitude of ascending node for that same orbit (fb) and therefore can use the same Equation (\ref{eq:orbital_binary_vx}), as for a single binary system.  Likewise, we can use Equation (\ref{eq:orbital_binary_sep}) for $\Delta z_\text{fb}$.

The barycentric transverse velocity, $v_\text{x,bary}$ is the velocity of the inner-binary (fb) caused by its motion within the outer-orbit (fbt) projected along the same direction as $v_\text{x,f,orb}$:

\begin{equation}
\begin{split}
    v_\text{x, bary} = \sqrt{\frac{G}{M_\text{fbt} a_{fbt} (1-e_{fbt}^2)}} m_t \Biggl[ &-\sin (\nu_{fb} (t)) \cos \omega_{fbt} \\
    &+ \left[ \cos (\nu_{fb} (t)) +  e_{fbt} \right] \sin \omega_{fbt} \Biggr]  \cos (\Omega_{fbt}-\Omega_{fb}) \>.
\end{split}
\end{equation}

\bibliographystyle{apj}
\bibliography{main}

\end{document}